\documentclass[
]
{amsart}
\usepackage{amsmath,amssymb,amsthm,eucal,amscd}

\newtheorem{thm}{Theorem}[section]

\newtheorem{lem}[thm]{Lemma}
\newtheorem{nota}[thm]{Notation}
\newtheorem{prop}[thm]{Proposition}

\newtheorem{hyp}[thm]{Hypothesis}

\theoremstyle{definition}
\newtheorem{df}[thm]{Definition}

\newtheorem{rem}[thm]{Remark}

\numberwithin{equation}{section}

\begin{document}

\def\<{\langle}
\def\>{\rangle}
\def\({\<\<}
\def\){\>\>}
\def\tri{| \hskip-0.02in |\hskip-0.02in |}
\def\1{{{1\hskip-2.0pt \rule{3pt}{0.2pt}\hskip-2.0pt
\rule{0.7pt}{6.6pt}
\hskip-2.2pt{\raise6.8pt\hbox{\rule{3pt}{0.2pt}}}\hskip0.8pt}}}
\def\1{{{{\Large \text{1}}\hskip-2.5pt \rule{3pt}{0.2pt}\hskip-2.0pt
\rule{0.7pt}{7.5pt}
\hskip-2.2pt{\raise7.6pt\hbox{\rule{3pt}{0.2pt}}}\hskip0.8pt}}}
\def\lnorm {\left|}
\def\rnorm {\right|}
\def\ad{\mathop{\rm ad}\nolimits}

\def\a{\mathfrak{A}}
\def\A{\mathcal{A}}
\def\CC{\mathbb{C}}
\def\C{\mathcal C}
\def\D{\mathcal{D}}
\def\DS{\rm{Diff}(S^1)}
\def\ds{\rm{diff}(S^1)}
\def\dsf{\rm{diff_0}(S^1)}
\def\e{\varepsilon}
\def\f{\mathfrak{F}}
\def\F{\mathcal{F}}
\def\g{\mathfrak g}
\def\G{\g_{\infty}}
\def\GG{G_{\infty}}
\def\h{\mathfrak h}
\def\H{\mathcal H}
\def\J{\mathcal J}
\def\k{\kappa}
\def\l{\lambda}
\def\m{\mathfrak m}
\def\M{\widetilde{M}}
\def\n{\nabla}
\def\nn{\tilde{\n}}
\def\o{\omega}
\def\O{\Omega}
\def\P{\mathfrak{P}}
\def\p{\Phi}
\def\r{\vec{r}}
\def\R{\mathbb{R}}
\def\RR{\widetilde{R}}
\def\Ric{\textrm{Ric}}
\def\s{\vec{s}}
\def\t{\theta}
\def\T{\widetilde{T}}
\def\Vir{\mathcal{V}_{c,h}}
\def\X{\mathbf{X}}
\def\y{\delta}
\def\z{\frac}

\def\so{\mathfrak{s}\mathfrak{o}_{HS}}
\def\gl{\mathfrak{g}\mathfrak{l}_{HS}}

\def\CM{G_{CM}}

\def\ll#1{\label{#1}
}

\def\zoxit#1
{\leavevmode\vbox{\hbox to 0pt{\hss
\raise1.8ex
\vbox to 0pt
{\vss\hrule\hbox
{\vrule\kern.75pt\vbox{\kern.75pt
\hbox{\tiny #1}
\kern.75pt}\kern.75pt\vrule}\hrule}}}}
\def\zb#1#2{\begin{#1}\ll{#2} 
}

\title[Virasoro algebra]
{Riemannian geometry of ${\rm Diff}(S^1)/S^1$ and representations of the Virasoro algebra}

\author[M. Gordina]{Maria Gordina}
\address[Maria Gordina]
{Department of Mathematics\\
University of Connecticut\\ 
Storrs, CT 06269, U.S.A.}
\thanks{
The research of the first author is partially supported by the NSF Grant DMS-0306468 and the Humboldt Foundation Research Fellowship.
}
\email{gordina@math.uconn.edu}
\author[P. Lescot]{Paul Lescot}
\address[Paul Lescot]{
INSSET--Universit\'{e} de Picardie \\
48 Rue Raspail \\
02100 Saint--Quentin, France\\
and LAMFA \\
Facult\'{e} de Math\'{e}matiques et Informatique \\
Universit\'{e} de Picardie \\
33 Rue Saint-Leu \\
80039 Amiens C\'{e}dex, France}

\email{paul.lescot@u-picardie.fr}



\date{\today}


\begin{abstract} The main result of the paper is a computation of the Ricci curvature of $\DS/S^1$. Unlike earlier results on the subject, we do not use the K\"{a}hler structure symmetries to compute the Ricci curvature, but rather rely on classical finite-dimensional results of Nomizu et al on Riemannian geometry of homogeneous spaces.  
\end{abstract}


\maketitle

\renewcommand{\contentsname}{Table of Contents}

\tableofcontents

\section{Introduction}\ll{S:INTRO}

Let $\DS$ be the group of orientation-preserving diffeomorphisms of the unit circle. This group is known as the Virasoro group in string theory. Then the quotient space $\DS/S^1$ describes those diffeomorphisms that fix a point on the circle. The geometry of this infinite-dimensional space has been of interest to physicists for a long time in connection with string theory and string field theory (e.g. \cite{BR87}, \cite{BR87-1}, \cite{Zumino87}). A. A. Kirillov and D. V. Yur'ev \cite{KirYu87} showed that the homogeneous space ${\rm Diff}(S^1)/S^1$ admits a left-invariant complex structure and can be canonically identified with $\mathcal{M}$, a certain space of univalent functions on the unit disk in $\CC$.

Our motivation comes from stochastic analysis on infinite-dimensional manifolds. In a series of papers written by H. Airault, V. Bogachev, P. Malliavin, A. Thalmeier (\cite{AB03, AM01, AMT03, AMT04}), the authors explored several possible approaches to the problem. For example, \cite{AMT04} is a first step in an attempt to construct a Brownian motion on $\mathcal{J}^{\infty}$, the space of smooth Jordan curves of the complex plane which can be described as the double quotient ${\rm {\rm SU}}(1,1)\backslash \DS /{\rm SU}(1,1)$. The connection between $\mathcal{J}^{\infty}$ and $\DS$ is given by the conformal welding. It is well-known that the behavior of a Brownian motion on a curved space (finite- or infinite-dimensional) is related to the geometry of this space. In particular, the lower bound of the Ricci curvature controls the growth of the Brownian motion, so it seems that a better understanding of geometry of ${\rm Diff}(S^1)/S^1$ might help in studying a Brownian motion on this homogeneous space.

The approach taken in \cite{BR87, BR87-1, Zumino87, KirYu87} is to describe the space ${\rm Diff}(S^1)/S^1$ as an infinite dimensional complex manifold with a K\"{a}hler metric, find the Riemann tensor corresponding to the K\"{a}hler structure, and finally compute the Ricci tensor. These computations use symmetries of the curvature tensor coming from the K\"{a}hler structure which are assumed to carry over from finite dimensions to infinite dimensions.

The aim of present article is to compute the Riemannian curvature tensor and the Ricci tensor for this space without using the K\"{a}hler structure. Rather we follow the path taken by the first author in \cite{Gor05}. There the Riemannian curvature tensor and the Ricci tensor were computed for a class of infinite-dimensional groups by using finite-dimensional computations of the Riemannian curvature tensor by J. Milnor in \cite{Milnor76} as definitions.

We will use the classical finite-dimensional results of K.Nomizu in \cite{Nom55} for homogeneous spaces as our definitions of basic geometric notions in this infinite-dimensional setting. The Virasoro algebra has a natural almost complex structure which has a zero torsion. This is why we can treat this structure as complex. Then using finite-dimensional methods we can find a covariant derivative compatible with the complex structure. For this covariant derivative we compute the corresponding Riemannian curvature tensor and the Ricci curvature. The main result of the paper is Theorem \ref{t:Ricci}. 

Finally we would like to mention that the main motivation for our work is a better understanding of a Brownian motion in ${\rm Diff}(S^1)/S^1$. There were numerous works exploring the connections between stochastic analysis and Riemannian geometry in infinite dimensions, mostly in loop groups and their extensions such as current groups, path spaces and complex Wiener spaces (e.g. \cite{Dr92}, \cite{ShigTani96}, \cite{Tani96}). 
 
\subsection*{Acknowledgement.} The authors would like to thank Bielefeld University, SFB 701 and the ZIF for invitations to Bielefeld in the summers of 2004 and 2005, during which most of the work on this paper was completed. Professor Michael R\"{o}ckner first suggested the topic, and provided encouragement, help and fruitful comments along the way. We are also grateful to Laurence Maillard-Teyssier for useful discussions concerning Riemannian geometry. The first author thanks the Humboldt Foundation for financial support of her stay in Germany in summer of 2005.

\section{Virasoro algebra}\ll{S:Vir}

In this exposition we follow \cite{AM01}.

\begin{nota}
Let $\DS$ be the group of orientation preserving $C^{\infty}$-diffeomorphisms of the unit circle, and $\ds$ its Lie algebra. The elements of $\ds$ will be identified with the left-invariant vector fields $f(t)\frac{d}{dt}$, with the Lie bracket given by

\[
[f,g]=f\dot{g}-\dot{f}g, f,g \in \ds.
\]
\end{nota}
The Lie algebra $\ds$ has a natural basis

\begin{equation}\ll{e:basis}
f_k=\cos{kt}, g_m=\sin{mt}, \ k=0, 1, 2..., m=1, 2....
\end{equation}
The Lie bracket in this basis satisfies the following identities

\begin{multline}\ll{e:comm}
[f_m, f_n]=\frac{1}{2}\left((m-n)g_{m+n}+(m+n)\frac{m-n}{|m-n|}g_{|m-n|}\right), \ m\not=n, \\
[g_m, g_n]=\frac{1}{2}\left((n-m)g_{m+n}+(m+n)\frac{m-n}{|m-n|}g_{|m-n|}\right), \ m\not=n, \\
[f_m, g_n]=\frac{1}{2}\left((n-m)f_{m+n}+(m+n)f_{|m-n|}\right).
\end{multline}

\begin{df} Suppose $c, h$ are positive constants. Then the Virasoro algebra $\Vir$ is the vector space $\R \oplus \ds$ with the Lie bracket given by

\begin{equation}\ll{e:Vbasis}
[a\k+f, b\k+g]_{\Vir}=\omega_{c,h}(f,g)\k+[f, g],
\end{equation}
where $k$ is the central element, and $\omega$ is the bilinear symmetric form

\[
\omega_{c,h}(f,g)=\int_{0}^{2\pi}\left((2h-\frac{c}{12})f^{'}(t)-\frac{c}{12}f^{(3)}(t)\right)g(t)\frac{dt}{2\pi}.
\]
\end{df}

\begin{rem}
If $h=0$, $c=6$, then $\omega_{c,h}$ is the fundamental cocycle $\omega$ (see \cite{AM01})

\[
\omega(f,g)=-\int_{0}^{2\pi} \left(f^{\prime}+f^{(3)}\right)g \frac{dt}{4\pi}.
\]
\end{rem}

\begin{rem}
A simple verification shows that $\omega_{c,h}$ satisfies the Jacobi identity, and therefore $\Vir$ is indeed a Lie algebra.

\end{rem}

\begin{nota} By $\dsf$ we denote the space of functions having mean $0$. This can be viewed as $\ds \slash S^1$, where $S^1$ is identified with constant vector fields corresponding to rotations of $S^1$.
\end{nota}

Then any element of $f \in \dsf$ can be written

\[
f(t)=\sum_{k=1}^{\infty} \left(a_k f_k+ b_k g_k\right).
\]
There is a natural endomorphism $J$ of $\dsf$ such that $J^2=-I$, namely,

\begin{equation}\ll{E:conv}
J(f)(t)=\sum_{k=1}^{\infty} \left(b_k f_k-a_k g_k\right).
\end{equation}

\begin{nota} For any $k \in \mathbb{Z}$
\[
\t_k=2hk+\frac{c}{12}(k^3-k).
\]
\end{nota}
Note that $\t_{-k}=-\t_k$, for any $k \in \mathbb{Z}$.
Let $b_0=0$, then

\begin{multline*}
\omega_{c,h}(f,Jf)=\int_{0}^{2\pi}\left((2h-\frac{c}{12})f^{'}(t)-\frac{c}{12}f^{(3)}(t)\right)(Jf)(t)\frac{dt}{2\pi}=\\
\int_{0}^{2\pi}\left(\sum_{k=0}^{\infty}\t_k(b_kf_k-a_kg_k)\right)\left(\sum_{m=1}^{\infty} b_m f_m-a_m g_m\right)\frac{dt}{2\pi}=
\frac{1}{2}\sum_{k=0}^{\infty}\t_k (a_k^2+b_k^2).
\end{multline*}
Thus $\omega_{c,h}$ induces a nondegenerate positive definite bilinear form on $\dsf$ if $h>0$.

\section{Riemannian geometry of ${\rm Diff}(S^1)/S^1$: definitions and preliminaries}\ll{S:geometry}

We use the finite-dimensional results of \cite{Nom55} as our definitions.
\begin{hyp}\ll{H1}
Let $\g$ be an infinite-dimensional Lie algebra equipped with an inner product $(\cdot ,\cdot )$. We assume that $\g$ is complete. Suppose that there are two subspaces, $\m$ and $\h$, of $\g$ such that $\g=\m\oplus \h$ as vector spaces. We assume that $\h$ is a Lie subalgebra of $\g$, and that $[\h, \m]\subset \m$. Note that $\m$ is not assumed to be a Lie subalgebra of $\g$.
\end{hyp}
In our setting $\g=\ds$, $\m=\dsf$, $\h=f_0\R$. Note that Hypothesis \ref{H1} is satisfied since for any $n \in \mathbb N$

\[
[f_{0},f_{n}]= -ng_{n}\in \m, \ 
[g_{0},g_{n}]= nf_{n}\in \m. 
\]
Let $G=\DS$ with the associated Lie algebra $\ds$, the subgroup $H=S^{1}$ with the Lie algebra $\h \subset \g$, then $\m$ is a Lie algebra naturally associated with the quotient
$\DS/S^{1}$.
For any $g \in \g$ we denote by $g_{\m}$ (respectively $g_{\h}$) its $\m$-(respectively $\h$-)component, that is, $g=g_{\m}+g_{\h}$, $g_{\m}\in \m$, $g_{\h}\in \h$. By Hypothesis \ref{H1} for any $h \in \h$ the adjoint representation $ad(h)=[h, \cdot]:\g \to \g$ maps $\m$ into $\m$. We will abuse notation by using $ad(h)$ for the corresponding endomorphism of $\m$.

Define

\[
B(f, g)=\omega_{c,h}(f, Jg)=\omega_{c,h}(g, Jf).
\]

\begin{prop}\ll{p:Binner}
$\<f, g\>=B(f, g)$ is an inner product on $diff_{0}(S^{1})$.
\end{prop}

\begin{proof} Follows from properties of $\omega_{c,h}$. In particular, for any $f \in \dsf$

\[
B(f, f)=\frac{1}{2}\sum_{k=1}^{\infty} \theta_k( a_k(f)^2+b_k(f)^2).
\]

\end{proof}

\begin{nota}\ll{n:B}The affine connection is defined by

\[
\alpha(x, y)=\frac{1}{2}[x,y]_{\m}+U(x,y),
\]
where $U$ is defined by

\[
B(U(x,y), z)=\frac{1}{2}\left(B([x,z]_{\m}, y)+B(x, [y,z]_{\m})\right)
\]
for any $x, y, z \in \m$. The relation between the covariant derivative $\n:\m \to End(\m)$ and $\alpha$ is given by

\[
\n_x y=\alpha(x, y)=\frac{1}{2}[x,y]_{\m}+U(x,y).
\]
\end{nota}

\begin{lem}\ll{L:lambda}
Let $\l_{m, n}=\frac{(2n+m)\t_m}{2\t_{m+n}}$ for any $n, m \in \mathbb{Z}$. Then

\[
\l_{m, n}=\l_{n, m}+\frac{m-n}{2}
\]

\end{lem}

\begin{proof}

\begin{multline*}
\l_{m, n}-\l_{n, m}=\frac{(2n+m)\t_m-(2m+n)\t_n}{\t_{m+n}}=
\\
\frac{2hm(2n+m)+\frac{c}{12}(m^3-m)(2n+m)-2hn(2m+n)+\frac{c}{12}(n^3-n)(2m+n)}{\t_{m+n}}
=\\
\frac{2h(m-n)(n+m)+\frac{c}{12}(m-n)((n+m)^3-(n+m))}{\t_{m+n}}=\frac{m-n}{2}.
\end{multline*}
\end{proof}

\begin{prop}\ll{P:Us} 

\begin{align*}
U(f_m,f_n)&=\frac{1}{2}\left[
(\l_{n, m}+\l_{m,n})g_{n+m}+(\l_{n, -m}-\l_{-m, n})g_{n-m}
\right]
\\
=&
\frac{1}{2}\left[
(\l_{n, m}+\l_{m,n})g_{n+m}+\frac{n+m}{2}g_{n-m}\right], n>m,
\\
\\
U(f_m,f_n)&=\frac{1}{2}\left[
(\l_{n, m}+\l_{m,n})g_{n+m}
+(\l_{n, -m}-\l_{-m, n})g_{m-n}
\right]
\\
&=
\frac{1}{2}\left[
(\l_{n, m}+\l_{m,n})g_{n+m}
+\frac{n+m}{2}g_{m-n}\right], m>n,
\\
\\
U(f_n,f_n)&=
\l_{n, n}g_{2n}
;
\end{align*}

\begin{align*}
U(f_m,g_n)&=\frac{1}{2}\left[
(\l_{-m, n}-\l_{n, -m})f_{n-m}
-(\l_{n, m}+\l_{m,n})f_{n+m}
\right]
\\
&=
\frac{1}{2}\left[
-\frac{n+m}{2}f_{n-m}
-(\l_{n, m}+\l_{m,n})f_{n+m}
\right], n>m,
\\
\\
U(f_m,g_n)&=\frac{1}{2}\left[
(\l_{n, -m}-\l_{-m, n})f_{m-n}
-(\l_{n, m}+\l_{m,n})f_{n+m}
\right]
\\
&=
\frac{1}{2}\left[
\frac{n+m}{2}f_{m-n}
-(\l_{n, m}+\l_{m,n})f_{n+m}
\right], m>n,
\\
\\
U(f_n,g_n)&=
-\l_{n, n}f_{2n}
;
\end{align*}

\begin{align*}
U(g_m,g_n)&=\frac{1}{2}\left[
(\l_{n, -m}-\l_{-m, n})g_{n-m}
-(\l_{n, m}+\l_{m,n})g_{n+m}
\right]
\\
&=
\frac{1}{2}\left[
\frac{n+m}{2}g_{n-m}
-(\l_{n, m}+\l_{m,n})g_{n+m}
\right], n>m,
\\
\\
U(g_m,g_n)&=\frac{1}{2}\left[
(\l_{n, -m}-\l_{-m, n})g_{m-n}
-(\l_{n, m}+\l_{m,n})g_{n+m}
\right],
\\
&=
\frac{1}{2}\left[
\frac{n+m}{2}g_{m-n}
-(\l_{n, m}+\l_{m,n})g_{n+m}
\right], m>n,
\\
\\
U(g_n,g_n)&=
-\l_{n, n}g_{2n}.
\end{align*}

\end{prop}

\begin{proof}

First,

\begin{multline*}
\omega_{c,h}(f_m, f_n)=-\int_0^{2\pi}\t_mg_mf_n\frac{dt}{2\pi}=0,\\
\omega_{c,h}(f_m, g_n)=-\int_0^{2\pi}\t_mg_mg_n\frac{dt}{2\pi}=-\frac{1}{2}\t_m\delta_{m, n},\\
\omega_{c,h}(g_m, f_n)=\int_0^{2\pi}\t_mf_mf_n\frac{dt}{2\pi}=\frac{1}{2}\t_m \delta_{m, n},\\
\omega_{c,h}(g_m, g_n)=\int_0^{2\pi}\t_mf_mg_n\frac{dt}{2\pi}=0,
\end{multline*}
and therefore

\[
B(f_m, f_n)=\frac{1}{2}\t_m \delta_{m, n},
B(f_m, g_n)=0,
B(g_m, f_n)=0, 
B(g_m, g_n)=\frac{1}{2}\t_m \delta_{m, n}.
\]

By the commutation relations \eqref{e:comm}

\begin{multline*}
B(U(f_m,f_n), f_k)=\frac{1}{2}\left(B([f_m,f_k]_{\m}, f_n)+B(f_m, [f_n,f_k]_{\m})\right)
\\
\frac{1}{4}\left[B((m-k)g_{m+k}+(m+k)\frac{m-k}{|m-k|}g_{|m-k|}, f_n)+B(f_m, (n-k)g_{n+k}+(n+k)\frac{n-k}{|n-k|}g_{|n-k|})\right]
\\
=0.
\end{multline*}

\begin{multline*}
B(U(f_m,f_n), g_k)=\frac{1}{2}\left(B([f_m,g_k]_{\m}, f_n)+B(f_m, [f_n,g_k]_{\m})\right)=
\\
=\frac{1}{4}\left[B(
(k-m)f_{m+k}+(m+k)f_{|m-k|}, f_n)+B(f_m, (k-n)f_{n+k}+(n+k)f_{|n-k|})\right]
=\\
\frac{1}{4}\left[
(k-m)B(f_{m+k}, f_n)+
(m+k)B(f_{|m-k|},f_n)+(k-n)B(f_m, f_{n+k})+(n+k)B(f_m, f_{|n-k|})\right]
=\\
\frac{1}{8}\left[
(k-m)\t_n\delta_{m+k, n}+
(m+k)\t_n\delta_{|m-k|, n}+(k-n)\t_m\delta_{m, n+k}+(n+k)\t_m\delta_{|n-k|, m}\right]
=\\
\frac{1}{8}[
(n-2m)\t_n\delta_{k, n-m}
+(2m+n)\t_n\delta_{k, m+n}
+(2m-n)\t_n\delta_{k, m-n}
+
\\(m-2n)\t_m\delta_{k, m-n}
+(2n+m)\t_m\delta_{k, n+m}
+(2n-m)\t_m\delta_{k, n-m}
]
=\\
\frac{1}{8}[
((n-2m)\t_n+(2n-m)\t_m)\delta_{k, n-m}
+((2m+n)\t_n+(2n+m)\t_m)\delta_{k, m+n}
+((2m-n)\t_n+(m-2n)\t_m)\delta_{k, m-n}
],
\end{multline*}
with the assumption that all the indices are positive. 
Thus

\begin{multline*}
U(f_m,f_n)=\frac{1}{4}\left[
\frac{(n-2m)\t_n+(2n-m)\t_m}{\t_{n-m}}g_{n-m}
+\frac{(2m+n)\t_n+(2n+m)\t_m}{\t_{n+m}}g_{n+m}
\right], n>m,
\\
U(f_m,f_n)=\frac{1}{4}\left[
\frac{(2m+n)\t_n+(2n+m)\t_m}{\t_{n+m}}g_{n+m}
+\frac{(2m-n)\t_n+(m-2n)\t_m}{\t_{m-n}}g_{m-n}
\right], m>n,
\\
U(f_n,f_n)=
\frac{3n\t_n}{2\t_{2n}}g_{2n}.
\end{multline*}

\begin{multline*}
B(U(f_m,g_n), f_k)=\frac{1}{2}\left(B([f_m,f_k]_{\m}, g_n)+B(f_m, [g_n,f_k]_{\m})\right)=
\\
\frac{1}{4}\left(B((m-k)g_{m+k}+(m+k)\frac{m-k}{|m-k|}g_{|m-k|}, g_n)-B(f_m, (n-k)f_{k+n}+(k+n)f_{|k-n|})\right)=
\\
\frac{1}{4}\left((m-k)B(g_{m+k}, g_n)
+(m+k)\frac{m-k}{|m-k|}B(g_{|m-k|}, g_n)
-(n-k)B(f_m, f_{k+n})
-(k+n)B(f_m, f_{|k-n|})\right)=
\\
\frac{1}{8}\left(
(m-k)\t_n\delta_{k, n-m}
+(m+k)\frac{m-k}{|m-k|}\t_n\delta_{n, |m-k|}
-(n-k)\t_m\delta_{k, m-n}
-(k+n)\t_m\delta_{m, |k-n|}
\right)=
\\
\frac{1}{8}\left(
((2m-n)\t_n-(2n-m)\t_m)\delta_{k, n-m}
+
(-(2n-m)\t_m+(2m-n)\t_n)\delta_{k, m-n}
\right.
\\
\left.
+(-(2m+n)\t_n-(m+2n)\t_m)\delta_{k, m+n}
\right).
\end{multline*}

\begin{multline*}
B(U(f_m,g_n), g_k)=\frac{1}{2}\left(B([f_m,g_k]_{\m}, g_n)+B(f_m, [g_n,g_k]_{\m})\right)=
\\
\frac{1}{4}
\left(B((k-m)f_{m+k}+(m+k)f_{|m-k|}, g_n)-B(f_m, (n-k)g_{k+n}+(k+n)\frac{k-n}{|k-n|}g_{|k-n|})\right)=0.
\end{multline*}
Thus

\begin{multline*}
U(f_m,g_n)=\frac{1}{4}\left[
\frac{(2m-n)\t_n-(2n-m)\t_m}{\t_{n-m}}f_{n-m}
-\frac{(2m+n)\t_n+(m+2n)\t_m)}{\t_{n+m}}f_{n+m}
\right], n>m,
\\
U(f_m,g_n)=\frac{1}{4}\left[
-\frac{(2m+n)\t_n+(m+2n)\t_m)}{\t_{n+m}}f_{n+m}
+\frac{(2m-n)\t_n-(2n-m)\t_m}{\t_{m-n}}f_{m-n}
\right], m>n,
\\
U(f_n,g_n)=
-\frac{3n\t_n}{2\t_{2n}}f_{2n}.
\end{multline*}

\begin{multline*}
B(U(g_m,g_n), f_k)=\frac{1}{2}\left(B([g_m,f_k]_{\m}, g_n)+B(g_m, [g_n,f_k]_{\m})\right)=
\\
\frac{1}{4}
\left(-B((m-k)f_{k+m}+(k+m)f_{|k-m|}, g_n)-B(g_m, (n-k)f_{k+n}+(k+n)f_{|k-n|})\right)=
\\
\frac{1}{4}
\left(
-(m-k)B(f_{k+m}, g_n)
-(k+m)B(f_{|k-m|}, g_n)
-(n-k)B(g_m, f_{k+n})
-(k+n)B(g_m, f_{|k-n|})\right)=0.
\end{multline*}

\begin{multline*}
B(U(g_m,g_n), g_k)=\frac{1}{2}\left(B([g_m,g_k]_{\m}, g_n)+B(g_m, [g_n,g_k]_{\m})\right)=
\\
\frac{1}{4}
\left(B((k-m)g_{m+k}+(m+k)\frac{m-k}{|m-k|}g_{|m-k|}, g_n)+B(g_m, (k-n)g_{n+k}+(n+k)\frac{n-k}{|n-k|}g_{|n-k|})\right)=
\\
\frac{1}{4}
\left(
(k-m)B(g_{m+k}, g_n)+
(m+k)\frac{m-k}{|m-k|}B(g_{|m-k|}, g_n)+
(k-n)B(g_m, g_{n+k})+
(n+k)\frac{n-k}{|n-k|}B(g_m, g_{|n-k|})
\right)=
\\
\frac{1}{8}
\left(
(k-m)\t_n\delta_{k, n-m}+
(m+k)\frac{m-k}{|m-k|}\t_n\delta_{n, |m-k|}+
(k-n)\t_m\delta_{k, m-n}+
(n+k)\frac{n-k}{|n-k|}\t_m\delta_{m, |n-k|}
\right)=
\\
\frac{1}{8}
\left(
((n-2m)\t_n+(2n-m)\t_m)\delta_{k, n-m}
+((2m-n)\t_n+(m-2n)\t_m)\delta_{k, m-n}
\right.
\\
\left.
-((2m+n)\t_n+(2n+m)\t_m)\delta_{k, m+n}.
\right)
\end{multline*}

Thus

\begin{multline*}
U(g_m,g_n)=\frac{1}{4}\left[
\frac{(n-2m)\t_n+(2n-m)\t_m}{\t_{n-m}}g_{n-m}
-\frac{(2m+n)\t_n+(2n+m)\t_m}{\t_{n+m}}g_{n+m}
\right], n>m,
\\
U(g_m,g_n)=\frac{1}{4}\left[
-\frac{(2m+n)\t_n+(2n+m)\t_m}{\t_{n+m}}g_{n+m}
+\frac{(2m-n)\t_n+(m-2n)\t_m}{\t_{m-n}}g_{m-n}
\right], m>n,
\\
U(g_n,g_n)=
-\frac{3n\t_n}{2\t_{2n}}g_{2n}.
\end{multline*}

\end{proof}

\begin{prop}\ll{P:nabla}

\begin{align*}
\n_{f_m}f_n&=\l_{m,n}g_{n+m}, \ n>m \\
\n_{f_m}f_n&=\l_{m,n}g_{n+m}+\frac{n+m}{2}g_{m-n}, \ n<m \\
\n_{f_n}f_n&=\l_{n, n}g_{2n}, \\
\n_{f_m}g_n&=-\l_{m,n}f_{n+m}, \ n>m \\
\n_{f_m}g_n&=-\l_{m,n}f_{n+m}+\frac{m+n}{2}f_{m-n}, \ n<m \\
\n_{f_n}g_n&=-\l_{n, n}f_{2n},\\
\n_{g_n}f_m&=\n_{f_m}g_n+[g_n, f_m]_{\m}\\
\n_{g_n}f_m&=-\l_{n,m}f_{n+m}-\frac{m+n}{2}f_{n-m}, \ n>m \\
\n_{g_n}f_m&=-\l_{n,m}f_{n+m}, \ n<m \\
\n_{g_n}f_n&=\n_{f_n}g_n=-\l_{n, n}f_{2n},\\
\n_{g_m}g_n&=-\l_{m,n}g_{n+m}, \ n>m \\
\n_{g_m}g_n&=\frac{n+m}{2}g_{m-n}-\l_{m,n}g_{n+m}, \ n<m \\
\n_{g_n}g_n&=-\l_{n, n}g_{2n}.
\end{align*}

\end{prop}

\begin{proof}
Here we use the commutation relations \eqref{e:comm}, Proposition \ref{P:Us} and Lemma \ref{L:lambda}.

If $n>m$, 
\begin{multline*}
\n_{f_m}f_n=\frac{1}{2}
[f_m, f_n]_{\m}+U(f_m, f_n)=\\
\frac{1}{4}\left((m-n)g_{m+n}-(m+n)g_{n-m}\right)
+\frac{1}{2}\left[
(\l_{n, m}+\l_{m,n})g_{n+m}+\frac{n+m}{2}g_{n-m}\right]=\\
\frac{1}{4}\left[
((m-n)+2(\l_{n, m}+\l_{m,n}))\right]g_{n+m}
=\frac{1}{4}\left[
((m-n)+2(\frac{n-m}{2}+2\l_{m,n}))\right]g_{n+m}
\\
=\l_{m,n}g_{n+m};
\end{multline*}

If $m>n$,
\begin{multline*}
\n_{f_m}f_n=\frac{1}{2}[f_m, f_n]_{\m}+U(f_m, f_n)=\\
\frac{1}{4}\left((m-n)g_{m+n}+(m+n)g_{m-n}\right)+\frac{1}{2}\left[
(\l_{n, m}+\l_{m,n})g_{n+m}
+\frac{n+m}{2}g_{m-n}\right]=\\
\frac{n+m}{2}g_{m-n}
+\l_{m,n}g_{n+m};
\end{multline*}

\[
\n_{f_n}f_n=U(f_n, f_n)=
\l_{n, n}g_{2n};
\]

If $n>m$, 
\begin{multline*}
\n_{f_m}g_n=\frac{1}{2}[f_m, g_n]_{\m}+U(f_m, g_n)=\\
\frac{1}{4}\left((n-m)f_{m+n}+(m+n)f_{n-m}\right)+\frac{1}{2}\left[
-\frac{n+m}{2}f_{n-m}
-(\l_{n, m}+\l_{m,n})f_{n+m}
\right]=\\
-\l_{m,n}f_{n+m};
\end{multline*}

If $m>n$,
\begin{multline*}
\n_{f_m}g_n=\frac{1}{2}[f_m, g_n]_{\m}+U(f_m, g_n)=\\
\frac{1}{4}\left((n-m)f_{m+n}+(m+n)f_{m-n}\right)+\frac{1}{2}\left[
\frac{n+m}{2}f_{m-n}
-(\l_{n, m}+\l_{m,n})f_{n+m}
\right]=
\\
-\l_{m,n}f_{n+m}
+\frac{m+n}{2}f_{m-n}
;
\end{multline*}

\[
\n_{f_n}g_n=\frac{1}{2}[f_n, g_n]_{\m}+U(f_n, g_n)=
-\l_{n, n}f_{2n}
;
\]

If $n>m$, 
\begin{multline*}
\n_{g_n}f_m=\frac{1}{2}[g_n, f_m]_{\m}+U(g_n, f_m)=\\
-\frac{1}{4}\left((n-m)f_{m+n}+(m+n)f_{n-m}\right)+\frac{1}{2}\left[
-\frac{n+m}{2}f_{n-m}
-(\l_{n, m}+\l_{m,n})f_{n+m}
\right]=\\
-\frac{1}{4}(n-m)f_{m+n}
-\frac{n+m}{2}f_{n-m}
-\frac{\l_{n, m}+\l_{m,n}}{2}f_{n+m}
=\\
-\frac{1}{4}(n-m)f_{m+n}
-\frac{n+m}{2}f_{n-m}
-\frac{\frac{n-m}{2}+2\l_{m,n}}{2}f_{n+m}
=\\
-\frac{n+m}{2}f_{n-m}
-\frac{n-m}{2}f_{m+n}
-\l_{m,n}f_{n+m}
=\\
\n_{f_m}g_n+[g_n, f_m].
\end{multline*}

If $m>n$,
\begin{multline*}
\n_{g_n}f_m=\frac{1}{2}[g_n, f_m]_{\m}+U(g_n, f_m)=\\
-\frac{1}{4}\left((n-m)f_{m+n}+(m+n)f_{m-n}\right)+\frac{1}{2}\left[
\frac{n+m}{2}f_{m-n}
-(\l_{n, m}+\l_{m,n})f_{n+m}
\right]=
\\
-\frac{n-m}{2}f_{m+n}-\l_{m, n}f_{n+m}
=\n_{f_m}g_n+[g_n, f_m]
;
\end{multline*}

\[
\n_{f_n}g_n=\frac{1}{2}[f_n, g_n]_{\m}+U(f_n, g_n)=
-\l_{n, n}f_{2n}
;
\]

If $n>m$, 
\begin{multline*}
\n_{g_m}g_n=\frac{1}{2}[g_m, g_n]_{\m}+U(g_m, g_n)=\\
\frac{1}{4}\left((n-m)g_{m+n}-(m+n)g_{n-m}\right)+\frac{1}{2}\left[
\frac{n+m}{2}g_{n-m}
-(\l_{n, m}+\l_{m,n})g_{n+m}
\right]=\\
\frac{n-m}{4}g_{m+n}
-\frac{1}{2}(\l_{n, m}+\l_{m,n})g_{n+m}
=
-\l_{m,n}g_{n+m};
\end{multline*}

If $m>n$,
\begin{multline*}
\n_{g_m}g_n=\frac{1}{2}[g_m, g_n]_{\m}+U(g_m, g_n)=\\
\frac{1}{4}\left((n-m)g_{m+n}+(m+n)g_{m-n}\right)+\frac{1}{2}\left[
\frac{n+m}{2}g_{m-n}
-(\l_{n, m}+\l_{m,n})g_{n+m}
\right]=\\
\frac{n-m}{4}g_{m+n}
+\frac{n+m}{2}g_{m-n}
-\frac{1}{2}(\frac{n-m}{2}+2\l_{m,n})g_{n+m}
=\frac{n+m}{2}g_{m-n}
-\l_{m,n}g_{n+m};
\end{multline*}

\[
\n_{g_n}g_n=\frac{1}{2}[g_n, g_n]_{\m}+U(g_n, g_n)=
U(g_n, g_n)=-\l_{n, n}g_{2n}.
\]

\end{proof}

\section{${\rm Diff}(S^1)/S^1$ as a K\"{a}hler manifold}

The goal of this section is to introduce an almost complex structure on $\Vir$, and then show that it is actually complex for an appropriately chosen connection. 

Recall that  $J: \Vir \to \Vir$ is an endomorphism defined by \eqref{E:conv}, or equivalently, on the basis $\{f_m, g_n\}$, $m, n=0, 1,...$ by

\[
Jf_m=-g_m, \ Jg_n=f_n.
\]
The next result is an analogue of the Newlander-Nirenberg theorem in our setting. This statement also appears in \cite{Air05} on p.255 as was communicated to us by H.Airault after we submitted the present paper.

\begin{prop}\ll{P:NN} The Nijenhuis tensor $N$ (the torsion of the almost complex structure $J$) defined by
\[
N(X,Y) = 2\left([JX, JY]_{\m}-[X,Y]_{\m}-J([X, JY])_{\m}-J[JX, Y]_{\m}\right)
\] 
vanishes on $\m=\dsf$. Therefore $J$ is a complex structure. 
\end{prop}

\begin{proof}
If $m\not=n$, then by \eqref{e:comm}
\begin{multline*}
N(f_m,f_n) = 2\left([Jf_m, Jf_n]_{\m}-[f_m,f_n]_{\m}-J([f_m, Jf_n])_{\m}-J[Jf_m, f_n]_{\m}\right)
=\\
2\left([g_m, g_n]_{\m}-[f_m,f_n]_{\m}+J([f_m, g_n])_{\m}+J[g_m, f_n]_{\m}\right)
=\\
2\left(
(n-m)g_{m+n}
+(n-m)Jf_{m+n}
\right)=0.
\end{multline*}
Then use

\begin{align*}
N(JX,Y)&= 
2\left(-[X, JY]_{\m}+[JX,J(JY)]_{\m}-J([JX, JY])_{\m}-J[X, J(JY)]_{\m}\right)
\\
&=N(X, JY)
\\
N(JX,Y)&= 
2\left(-[X, JY]_{\m}+[JX,J(JY)]_{\m}-J([JX, JY])_{\m}-J[X, J(JY)]_{\m}\right)
=\\
& -2J\left(-J[X, JY]_{\m}-J[JX,Y]_{\m}+([JX, JY])_{\m}-[X, Y]_{\m}\right)
=-JN(X,Y)
\end{align*}

to see that

\[
N(f_m,f_n)=N(Jg_m,f_n)=N(g_m,Jf_n)=-N(g_m,g_n)=0,
\]
and
\[
N(f_m,g_n)=-N(g_n,f_m)=N(Jg_m, g_n)=-J N(g_m, g_n)=0.
\]

\end{proof}

\begin{lem}
$J$ is a complex structure on $\m=\dsf$ with the covariant derivative

\begin{align*}
(\n_{f_m} J) ({f_n})=&0, \ n\geqslant m,
\\
(\n_{f_m} J) ({f_n})=&-(m+n)f_{m-n}, \ n<m,
\\
(\n_{f_m} J) ({g_n})=&0, \ n\geqslant m,
\\
(\n_{f_m} J) ({g_n})=&(n+m)g_{m-n}, \ n<m,
\\
(\n_{g_m} J) ({f_n})=&0, \ n\geqslant m,
\\
(\n_{g_m} J) ({f_n})=&-(n+m)g_{m-n}, \ n<m,
\\
(\n_{g_m} J) ({g_n})=& 0, \ n\geqslant m,
\\
(\n_{g_m} J) ({g_n})=&-(m+n)f_{m-n}, \ n<m.
\end{align*}

\end{lem}

\begin{proof}  We will use the fact that

\[
(\n_x J) (y)=\n_x (Jy)-J(\n_x y).
\]

If $n>m$, then

\begin{multline*}
(\n_{f_m} J) ({f_n})=-\n_{f_m} g_n-J(\n_{f_m} {f_n})
=\\
\l_{m,n}f_{n+m}-\l_{m,n}J(g_{n+m})
=
\l_{m,n}f_{n+m}-\l_{m,n}f_{n+m}=0
\end{multline*}

If $n<m$, then

\begin{multline*}
(\n_{f_m} J) ({f_n})=-\n_{f_m} g_n-J(\n_{f_m} {f_n})
=\\
\l_{m,n}f_{n+m}-\frac{m+n}{2}f_{m-n}-\l_{m,n}J(g_{n+m})-\frac{n+m}{2}J(g_{m-n})
=\\
\l_{m,n}f_{n+m}
-\frac{m+n}{2}f_{m-n}
-\l_{m,n}f_{n+m}
-\frac{n+m}{2}f_{m-n}
=
-(m+n)f_{m-n}
\end{multline*}

\begin{multline*}
(\n_{f_n} J) ({f_n})=-\n_{f_n} g_n-J(\n_{f_n} {f_n})
=\\
\l_{n, n}f_{2n}-\l_{n, n}J(g_{2n})=\l_{n, n}f_{2n}-\l_{n, n}f_{2n}=0
\end{multline*}

If $n>m$, then

\begin{multline*}
(\n_{f_m} J) ({g_n})=\n_{f_m} f_n-J(\n_{f_m} {g_n})
=\\
\l_{m,n}g_{n+m}+\l_{m,n}J(f_{n+m})
=\l_{m,n}g_{n+m}-\l_{m,n}g_{n+m}=0
\end{multline*}

If $n<m$, then

\begin{multline*}
(\n_{f_m} J) ({g_n})=\n_{f_m} f_n-J(\n_{f_m} {g_n})
=\\
\l_{m,n}g_{n+m}+\frac{n+m}{2}g_{m-n}+\l_{m,n}J(f_{n+m})-\frac{m+n}{2}J(f_{m-n})
=\\
\l_{m,n}g_{n+m}+\frac{n+m}{2}g_{m-n}-\l_{m,n}g_{n+m}+\frac{m+n}{2}g_{m-n}
=
(n+m)g_{m-n}
\end{multline*}

\begin{multline*}
(\n_{f_n} J) ({g_n})=\n_{f_n} f_n-J(\n_{f_n} {g_n})
=\\
\n_{f_n} f_n-J(\n_{f_n} {g_n})
=
\l_{n, n}g_{2n}+\l_{n, n}J(f_{2n})=0
\end{multline*}

If $n>m$, then

\begin{multline*}
(\n_{g_m} J) ({f_n})=-\n_{g_m} g_n-J(\n_{g_m} {f_n})
=\\
\l_{m,n}g_{n+m}+\l_{m,n}J(f_{n+m})
=
\l_{m,n}g_{n+m}-\l_{m,n}g_{n+m}=0
\end{multline*}

If $n<m$, then

\begin{multline*}
(\n_{g_m} J) ({f_n})=-\n_{g_m} g_n-J(\n_{g_m} {f_n})
=\\
-\frac{n+m}{2}g_{m-n}+\l_{m,n}g_{n+m}
+\l_{m,n}J(f_{n+m})+\frac{m+n}{2}J(f_{m-n})
=\\
-\frac{n+m}{2}g_{m-n}
-\frac{m+n}{2}g_{m-n}
+\l_{m,n}g_{n+m}
-\l_{m,n}g_{n+m}=
-(n+m)g_{m-n}
\end{multline*}

\begin{multline*}
(\n_{g_n} J) ({f_n})=-\n_{g_n} g_n-J(\n_{g_n} {f_n})
=\\
\l_{n, n}g_{2n}+\l_{n, n}J(f_{2n})
=
\l_{n, n}g_{2n}-\l_{n, n}g_{2n}=0
\end{multline*}

If $n>m$, then

\begin{multline*}
(\n_{g_m} J) ({g_n})=\n_{g_m} f_n-J(\n_{g_m} {g_n})
=\\
-\l_{m,n}f_{n+m}
+\l_{m,n}J(g_{n+m})
=-\l_{m,n}f_{n+m}+\l_{m,n}f_{n+m}=0
\end{multline*}

If $n<m$, then

\begin{multline*}
(\n_{g_m} J) ({g_n})=\n_{g_m} f_n-J(\n_{g_m} {g_n})
=\\
-\l_{m,n}f_{n+m}-\frac{m+n}{2}f_{m-n}-\frac{n+m}{2}J(g_{m-n})+\l_{m,n}J(g_{n+m})
=\\
-\l_{m,n}f_{n+m}-\frac{m+n}{2}f_{m-n}-\frac{n+m}{2}f_{m-n}+\l_{m,n}f_{n+m}
=
-(m+n)f_{m-n}
\end{multline*}

\begin{multline*}
(\n_{g_n} J) ({g_n})=\n_{g_n} f_n-J(\n_{g_n} {g_n})
=\\
-\l_{n, n}f_{2n}+\l_{n, n}J(g_{2n})
=
-\l_{n, n}f_{2n}+\l_{n, n}f_{2n}=0
\end{multline*}

\end{proof}

\begin{lem}\ll{L:Q} Let $Q$ be the tensor field of type $(1, 2)$ defined by
\[
4Q(x, y)=(\n_{Jy}J)x+J((\n_{y}J)x)+2J((\n_{x}J)y)
\]

Then

\begin{align*}
Q(f_m, f_n)=\frac{m+n}{2}g_{n-m}, \ n>m
\\
Q(f_m, f_n)=\frac{m+n}{2}g_{m-n}, / n<m
\\
Q(f_n, f_n)=0
\\
Q(f_m, g_n)=-\frac{m+n}{2}f_{n-m}, \ n>m
\\
Q(f_m, g_n)=\frac{m+n}{2}f_{m-n}, \ n<m
\\
Q(f_n, g_n)=0
\\
Q(g_m, f_n)=\frac{m+n}{2}f_{n-m}, \ n>m
\\
Q(g_m, f_n)=-\frac{m+n}{2}f_{m-n}, \ n<m
\\
Q(g_n, f_n)=0
\\
Q(g_m, g_n)=\frac{m+n}{2}g_{n-m}, \ n>m
\\
Q(g_m, g_n)=\frac{m+n}{2}g_{m-n}, \ n<m
\\
Q(g_n, g_n)=0
\end{align*}

\end{lem}

\begin{proof}

If $n>m$, then

\begin{multline*}
4Q(f_m, f_n)=(\n_{Jf_n}J)f_m+J((\n_{f_n}J)f_m)+2J((\n_{f_m}J)f_n)
=\\
(n+m)g_{n-m}-(m+n)J(f_{n-m})
=
2(m+n)g_{n-m}
\end{multline*}

If $n<m$, then

\begin{multline*}
4Q(f_m, f_n)=(\n_{Jf_n}J)f_m+J((\n_{f_n}J)f_m)+2J((\n_{f_m}J)f_n)
=\\
2(m+n)g_{m-n}
\end{multline*}

\begin{multline*}
4Q(f_n, f_n)=(\n_{Jf_n}J)f_n+J((\n_{f_n}J)f_n)+2J((\n_{f_n}J)f_n)
=\\
-(\n_{g_n}J)f_n+3J((\n_{f_n}J)f_n)=0
\end{multline*}

If $n>m$, then

\begin{multline*}
4Q(f_m, g_n)=(\n_{Jg_n}J)f_m+J((\n_{g_n}J)f_m)+2J((\n_{f_m}J)g_n)
=\\
-(m+n)f_{n-m}-(m+n)J(g_{n-m})
=
-2(m+n)f_{n-m}
\end{multline*}

If $n<m$, then

\begin{multline*}
4Q(f_m, g_n)=(\n_{Jg_n}J)f_m+J((\n_{g_n}J)f_m)+2J((\n_{f_m}J)g_n)
=\\
2(n+m)J(g_{m-n})=2(n+m)f_{m-n}
\end{multline*}

\[
4Q(f_n, g_n)=(\n_{Jg_n}J)f_n+J((\n_{g_n}J)f_n)+2J((\n_{f_n}J)g_n)=0
\]

If $n>m$, then

\begin{multline*}
4Q(g_m, f_n)=(\n_{Jf_n}J)g_m+J((\n_{f_n}J)g_m)+2J((\n_{g_m}J)f_n)
=\\
(m+n)f_{n-m}+(m+n)J(g_{n-m})=2(m+n)f_{n-m}
\end{multline*}

If $n<m$, then

\begin{multline*}
4Q(g_m, f_n)=(\n_{Jf_n}J)g_m+J((\n_{f_n}J)g_m)+2J((\n_{g_m}J)f_n)
=\\
-(n+m)2J(g_{m-n})=-(n+m)2f_{m-n}
\end{multline*}
\begin{multline*}
4Q(g_n, f_n)=(\n_{Jf_n}J)g_n+J((\n_{f_n}J)g_n)+2J((\n_{g_n}J)f_n)
=\\
-(\n_{g_n}J)g_n+J((\n_{f_n}J)g_n)+2J((\n_{g_n}J)f_n)=0
\end{multline*}

If $n>m$, then

\begin{multline*}
4Q(g_m, g_n)=(\n_{Jg_n}J)g_m+J((\n_{g_n}J)g_m)+2J((\n_{g_m}J)g_n)
=\\
(n+m)g_{n-m}-(m+n)J(f_{n-m})=2(n+m)g_{n-m}
\end{multline*}

If $n<m$, then

\begin{multline*}
4Q(g_m, g_n)=(\n_{Jg_n}J)g_m+J((\n_{g_n}J)g_m)+2J((\n_{g_m}J)g_n)
=\\
-2(m+n)J(f_{m-n})=2(m+n)g_{m-n}
\end{multline*}

\begin{multline*}
4Q(g_n, g_n)=(\n_{Jg_n}J)g_n+J((\n_{g_n}J)g_n)+2J((\n_{g_n}J)g_n)
=\\
(\n_{f_n}J)g_n+3J((\n_{g_n}J)g_n)=0
\end{multline*}

\end{proof}

\begin{df}
The new covariant derivative is defined by

\[
\nn_xy=\n_xy-Q(x, y).
\]
\end{df}
Then combining the results of Proposition \ref{P:nabla} and Lemma \ref{L:Q} we see that

\begin{align*}
\nn_{f_m}f_n=&\n_{f_m}f_n-Q(f_m, f_n)=\l_{m,n}g_{n+m}-\frac{m+n}{2}g_{n-m}, \ n>m
\\
\nn_{f_m}f_n=&\n_{f_m}f_n-Q(f_m, f_n)=\l_{m,n}g_{n+m}, \ n<m
\\
\nn_{f_n}f_n=&\n_{f_n}f_n-Q(f_n, f_n)=\l_{n, n}g_{2n}
\\
\nn_{f_m}g_n=&\n_{f_m}g_n-Q(f_m, g_n)=\frac{m+n}{2}f_{n-m}-\l_{m,n}f_{n+m}, \ n>m
\\
\nn_{f_m}g_n=&\n_{f_m}g_n-Q(f_m, g_n)=-\l_{m,n}f_{n+m}, \ n<m
\\
\nn_{f_n}g_n=&\n_{f_n}g_n-Q(f_n, g_n)=-\l_{n, n}f_{2n}
\\
\nn_{g_n}f_m=&\n_{g_n}f_m-Q(g_n, f_m)=-\l_{n,m}f_{n+m}, \ n>m
\\
\nn_{g_n}f_m=&\n_{g_n}f_m-Q(g_n, f_m)=-\l_{n,m}f_{n+m}-\frac{m+n}{2}f_{m-n}, \ n<m
\\
\nn_{g_n}f_n=&\n_{g_n}f_n-Q(g_n, f_n)=-\l_{n, n}f_{2n}
\\
\nn_{g_m}g_n=&\n_{g_m}g_n-Q(g_m, g_n)=-\l_{m,n}g_{n+m}-\frac{m+n}{2}g_{n-m}, \ n>m
\\
\nn_{g_m}g_n=&\n_{g_m}g_n-Q(g_m, g_n)=-\l_{m,n}g_{n+m}, \ n<m
\\
\nn_{g_n}g_n=&\n_{g_n}g_n-Q(g_n, g_n)=-\l_{n, n}g_{2n}
\end{align*}

\begin{thm}
The covariant derivative $\nn$ has the following properties

\begin{enumerate}
\item $\nn$ is the Levi-Civita covariant derivative, that is, it is metric compatible and torsion free;

\item $\nn$ is not a Hilbert-Schmidt operator.
\end{enumerate}
\end{thm}

\begin{rem} The original covariant derivative $\n$ is also torsion free, which can be checked by a direct computation 

\begin{align*}
T_{\nabla}(X,Y)
= & \nabla_{X}(Y)-\nabla_{Y}(X)-[X,Y]_{\m}
\\ 
= &\left(\frac{1}{2}[X,Y]_{\m}+U(X,Y)\right)-\left(\frac{1}{2}[Y,X]_{\m}+U(Y,X)\right)-[X,Y]_{\m}
\\ 
= & U(X,Y)-U(Y,X). 
\end{align*}
Note that  $U(X,Y)$ is symmetric in $(X,Y)$ due to the symmetry of $B$ as can be seen from Notation \ref{n:B}, and
therefore $T_{\nabla}=0$.
Similarly to the finite-dimensional case the new covariant derivative $\nn$ is torsion free if the almost complex structure $J$ has no torsion. This is indeed the case by Proposition \ref{P:NN}.
\end{rem}

\begin{proof}

\begin{enumerate}
\item

\[
\T(x,y)=T_{\nn}(x,y)=\nn_x y-\nn_y x-[x,y]_{\m}
\]
Let $m\not=n$, then

\begin{multline*}
\T(f_m,{f_n})=\nn_{f_m} {f_n}-\nn_{f_n} {f_m}-\frac{m-n}{2}g_{m+n}-\frac{m+n}{2}\frac{m-n}{|m-n|}g_{|m-n|}
=\\
\frac{m-n}{2}g_{m+n}
+\frac{m+n}{2}\frac{m-n}{|m-n|}g_{|m-n|}
-\frac{m-n}{2}g_{m+n}
-\frac{m+n}{2}\frac{m-n}{|m-n|}g_{|m-n|}
=0
\end{multline*}

\begin{multline*}
\T(f_m,{g_n})=\nn_{f_m} {g_n}-\nn_{g_n} {f_m}-\frac{n-m}{2}f_{m+n}-\frac{m+n}{2}f_{|m-n|}
=\\
\frac{m+n}{2}f_{|m-n|}+(\l_{n,m}-\l_{m,n})f_{n+m}-\frac{n-m}{2}f_{m+n}-\frac{m+n}{2}f_{|m-n|}
=0
\end{multline*}

\begin{multline*}
\T(g_m,{f_n})=\nn_{g_m} {f_n}-\nn_{f_n} {g_m}+\frac{m-n}{2}f_{m+n}+\frac{m+n}{2}f_{|m-n|}
=\\
-\frac{m+n}{2}f_{|m-n|}+(\l_{n,m}-\l_{m,n})f_{n+m}+\frac{m-n}{2}f_{m+n}+\frac{m+n}{2}f_{|m-n|}=0
\end{multline*}

\begin{multline*}
\T(g_m,{g_n})=\nn_{g_m} {g_n}-\nn_{g_n} {g_m}-\frac{n-m}{2}g_{m+n}-\frac{m+n}{2}\frac{m-n}{|m-n|}g_{|m-n|}
=\\
(\l_{n,m}-\l_{m,n})g_{n+m}+\frac{m+n}{2}\frac{m-n}{|m-n|}g_{|m-n|}-\frac{n-m}{2}g_{m+n}-\frac{m+n}{2}\frac{m-n}{|m-n|}g_{|m-n|}=0
\end{multline*}

\item

\begin{multline*}
\sum_{m=1}^{\infty}\left(
\left< \nn_{f_m}f_n, \nn_{f_m}f_n \right> +
\left< \nn_{g_m}f_n, \nn_{g_m}f_n \right>
\right)
=\\
\sum_{m=1}^{n-1}
\left(
\frac{\l_{m,n}^2\t_{n+m}}{\t_n\t_m}+\frac{(m+n)^2\t_{n-m}}{4\t_n\t_m}
\right)
+
\frac{\l_{n,n}^2\t_{2n}}{\t_n^2}
+\sum_{m=n+1}^{\infty}
\frac{\l_{m,n}^2\t_{n+m}}{\t_n\t_m}=+\infty. 
\end{multline*}

\end{enumerate}
\end{proof}

\begin{nota}
Let $n \in \mathbb N$, then define

\[
L_m=f_m+ig_m, \ L_{-m}=f_m-ig_m, i^2=-1.
\]
\end{nota}

\begin{lem}

\begin{align*}
[L_m, L_n]&=i(n-m)L_{m+n};\\
[L_{-m}, L_n]&=i(m+n)L_{n-m};\\
[L_{m}, L_{-n}]&=-i(m+n)L_{m-n};\\
[L_{-m}, L_{-n}]&=i(m-n)L_{-m-n}.
\end{align*}

\end{lem}

\begin{proof}
\begin{multline*}
[L_m, L_n]=[f_m, f_n]-[g_m, g_n]+i\left([f_m, g_n]+[g_m, f_n]\right)
=\\
(m-n)g_{m+n}+i(n-m)f_{m+n}=i(n-m)L_{m+n}
\end{multline*}

\begin{multline*}
[L_{-m}, L_n]=[f_m, f_n]+[g_m, g_n]+i\left([f_m, g_n]-[g_m, f_n]\right)
=\\
(m+n)\frac{m-n}{|m-n|}g_{|m-n|}+i(m+n)f_{|m-n|}=i(m+n)L_{n-m}
\end{multline*}

\begin{multline*}
[L_{m}, L_{-n}]=[f_m, f_n]+[g_m, g_n]+i\left(-[f_m, g_n]+[g_m, f_n]\right)
=\\
(m+n)\frac{m-n}{|m-n|}g_{|m-n|}-i(m+n)f_{|m-n|}=-i(m+n)L_{m-n}
\end{multline*}

\begin{multline*}
[L_{-m}, L_{-n}]=[f_m, f_n]-[g_m, g_n]-i\left([f_m, g_n]+[g_m, f_n]\right)
=\\
(m-n)g_{m+n}-i(n-m)f_{m+n}=i(m-n)L_{-m-n}
\end{multline*}
\end{proof}

\begin{lem}
\begin{align*}
\nn_{L_{m}} L_{n}&=-2i\l_{m, n}L_{m+n}
;\\
\nn_{L_{-m}} L_{n}&=i(m+n)L_{n-m}, \ n>m
;\\
\nn_{L_{-m}} L_{n}&=0, \ m>n
;\\
\nn_{L_{m}} L_{-n}&=-i(m+n)L_{m-n}, \ n>m
;\\
\nn_{L_{m}} L_{-n}&=0, \ m>n
;\\
\nn_{L_{-m}} L_{-n}&=2i\l_{m, n}L_{-m-n}
;\\
\nn_{L_{n}} L_{n}&=-2i\l_{n, n}L_{2n}
;\\
\nn_{L_{-n}} L_{n}&=0
;\\
\nn_{L_{n}} L_{-n}&=0
;\\
\nn_{L_{-n}} L_{-n}&=2i\l_{n, n}L_{-2n}.
\end{align*}

\end{lem}

\begin{proof}

\begin{multline*}
\nn_{L_{m}} L_{n}=
\nn_{f_m}f_n-\nn_{g_m} g_n
+i\left(\nn_{f_m} g_n+\nn_{g_m} f_n\right)
=\\
2\l_{m,n}g_{m+n}-2i\l_{m, n}f_{m+n}=
-2i\l_{m, n}L_{n+m}.
\end{multline*}

If $n>m$, then

\begin{multline*}
\nn_{L_{-m}} L_{n}=\nn_{f_m}f_n+\nn_{g_m} g_n+i\left(\nn_{f_m} g_n-\nn_{g_m}f_n\right)
=\\
-(m+n)g_{n-m}+i(m+n)f_{n-m}
=i(m+n)L_{n-m}
\end{multline*}

If $n<m$, then

\[
\nn_{L_{-m}} L_{n}=\nn_{f_m}f_n+\nn_{g_m} g_n+i\left(\nn_{f_m} g_n-\nn_{g_m}f_n\right) 
=0
\]

If $n>m$, then

\begin{multline*}
\nn_{L_{m}} L_{-n}=\nn_{f_m}f_n+\nn_{g_m} g_n-i\left(\nn_{f_m} g_n-\nn_{g_m}f_n\right)
=\\
-(m+n)g_{n-m}-i(m+n)f_{n-m}=-i(m+n)L_{m-n}
\end{multline*}

If $n<m$, then

\[
\nn_{L_{m}} L_{-n}=\nn_{f_m}f_n+\nn_{g_m} g_n-i\left(\nn_{f_m} g_n-\nn_{g_m}f_n\right)=0
\]

\begin{multline*}
\nn_{L_{-m}} L_{-n}=\nn_{f_m}f_n-\nn_{g_m}g_n-i\left(\nn_{f_m}g_n+\nn_{g_m}f_n\right)
=\\
2\l_{m,n}g_{m+n}+2i\l_{m, n}f_{m+n}=
2i\l_{m, n}L_{-m-n}.
\end{multline*}

\begin{multline*}
\nn_{L_{n}} L_{n}=
\nn_{f_n}f_n-\nn_{g_n} g_n
+i\left(\nn_{f_n} g_n+\nn_{g_n} f_n\right)
=\\
2\l_{n,n}g_{2n}-2i\l_{n,n}f_{2n}=-2i\l_{n,n}L_{2n}
\end{multline*}

\[
\nn_{L_{-n}} L_{n}=\nn_{f_n}f_n+\nn_{g_n} g_n+i\left(\nn_{f_n} g_n-\nn_{g_n}f_n\right) 
=0
\]

\[
\nn_{L_{n}} L_{-n}=\nn_{f_n}f_n+\nn_{g_n} g_n-i\left(\nn_{f_n} g_n-\nn_{g_n}f_n\right)=0
\]

\begin{multline*}
\nn_{L_{-n}} L_{-n}=\nn_{f_n}f_n-\nn_{g_n}g_n-i\left(\nn_{f_n}g_n+\nn_{g_n}f_n\right)
=\\
2\l_{n,n}g_{2n}+2i\l_{n,n}f_{2n}=
2i\l_{n, n}L_{-2n}.
\end{multline*}

\end{proof}

\begin{df} The curvature tensor is defined by
\[ 
\RR_{xy}=\nn_x\nn_y-\nn_y\nn_x-\nn_{[x,y]_{\m_{\CC}}}-ad([x,y]_{\h_{\CC}}), \ x,y \in \g;
\]
the Ricci tensor $\Ric(x, y)$ is then the trace of the map $z \mapsto \RR_{zx}y$. 

\end{df}

\begin{thm}\ll{t:Ricci} The only non-zero components of the Ricci tensor are

\[
\Ric(\frac{L_n}{\sqrt{\t_n}}, \frac{L_{-n}}{\sqrt{\t_n}})
=
-
\frac{13n^3-n}{6\t_n}, \ n\in \mathbb{Z}, n\not=0.
\]

\end{thm}

\begin{proof}
Note that for any $\alpha, \beta, \gamma \in \mathbb{Z}$ we have $\RR_{L_{\gamma} L_{\alpha}}L_{\beta}=C_{\alpha \beta \gamma}L_{\alpha+\beta+\gamma}$ for some $C_{\alpha \beta \gamma}\in \mathbb C$. Therefore the only non-zero components of $\Ric(L_{\alpha}, L_{\beta})$ are when $\alpha+\beta=0$. 

Suppose $m\not=n$, then

\begin{multline*}
\RR_{L_m, L_n}L_{-n}=
\nn_{L_m}\nn_{L_n}L_{-n}-\nn_{L_n}\nn_{L_m}L_{-n}-\nn_{[{L_m},{L_n}]_{\m}}L_{-n}-ad([{L_m},{L_n}]_{\h})L_{-n}
=\\
-\nn_{L_n}\nn_{L_m}L_{-n}-i(n-m)\nn_{L_{m+n}}L_{-n}
=
-\nn_{L_n}\nn_{L_m}L_{-n}
\end{multline*}

If $m>n$, then

\[
\RR_{L_m, L_n}L_{-n}=0.
\]
If $m<n$, then

\[
\RR_{L_m, L_n}L_{-n}=
i(m+n)\nn_{L_n}L_{m-n}=0.
\]

If $m\not=n$, then
\begin{multline*}
\RR_{L_{-m}, L_n}L_{-n}=
\nn_{L_{-m}}\nn_{L_n}L_{-n}-\nn_{L_n}\nn_{L_{-m}}L_{-n}-\nn_{[{L_{-m}},{L_n}]_{\m}}L_{-n}-ad([{L_{-m}},{L_n}]_{\h})L_{-n}
=\\
-\nn_{L_n}\nn_{L_{-m}}L_{-n}
-i(m+n)\nn_{L_{n-m}}L_{-n}
=
-2i\l_{m, n}\nn_{L_n}L_{-m-n}
-i(m+n)\nn_{L_{n-m}}L_{-n}
=\\
-2(m+2n)\l_{m, n}L_{-m}
-i(m+n)\nn_{L_{n-m}}L_{-n}
\end{multline*}

If $m>n$, then

\[
\RR_{L_{-m}, L_n}L_{-n}=
-2(m+2n)\l_{m, n}L_{-m}
+2(m+n)\l_{m-n, n}L_{-m}
\]

If $m<n$, then

\[
\RR_{L_{-m}, L_n}L_{-n}=
-2(m+2n)\l_{m, n}L_{-m}
-(2n-m)(m+n)L_{-m}
\]

\begin{multline*}
\RR_{L_{-n}, L_n}L_{-n}=
\nn_{L_{-n}}\nn_{L_n}L_{-n}-\nn_{L_n}\nn_{L_{-n}}L_{-n}-\nn_{[{L_{-n}},{L_n}]_{\m}}L_{-n}-ad([{L_{-n}},{L_n}]_{\h})L_{-n}
=\\
-6n\l_{n, n}L_{-n}
-2n^2L_{-n}
\end{multline*}

Thus
\begin{multline*}
\Ric(\frac{L_n}{\sqrt{\t_n}}, \frac{L_{-n}}{\sqrt{\t_n}})
=\\
-\sum_{m=1}^{n}\frac{(m+n)(2n-m)+2(m+2n)\l_{m,n}}{\t_n}
+\sum_{m=n+1}^{\infty}\frac{2(m+n)\l_{m-n, n}-2(m+2n)\l_{m, n}}{\t_n}
=\\
-\sum_{m=1}^{n}\frac{(m+n)(2n-m)+2(m+2n)\l_{m,n}}{\t_n}
+\sum_{m=1}^{n}\frac{2(m+2n)\l_{m, n}}{\t_n}
=\\
-\sum_{m=1}^{n}\frac{(m+n)(2n-m)}{\t_n}=
-
\frac{13n^3-n}{6\t_n}.
\end{multline*}

\end{proof}

\end{document}